\documentclass[preprint,12pt,authoryear]{elsarticle}

\usepackage{times}
\usepackage{hyperref}
\usepackage{url}
\usepackage{listings} 
\usepackage{todonotes}  
\usepackage{amsmath,amsthm,amssymb}

\usepackage{graphicx,epstopdf}
\usepackage{subcaption}

\newcommand{\ee}{\mathbf{E}}
\newcommand{\var}{\mathbf{Var}}

\newcommand{\argmax}{\operatornamewithlimits{argmax}}
\newcommand{\Ito}{It$\mathrm{\bar{o}}$~}

\begin{document}

\begin{frontmatter}
\title{ Efficient Computation of the Quasi Likelihood function for Discretely Observed Diffusion Processes}


\author[Josef]{Lars Josef~H\"{o}\"{o}k}
\ead{josef.hook@it.uu.se}

\author[Erik]{Erik Lindstr\"{o}m}

\address[Josef]{Division of Scientific Computing, Department of Information Technology,
Uppsala University, Box 337, SE-751 05 Uppsala, Sweden}
\address[Erik]{Mathematical Statistics, Centre for Mathematical Sciences, Lund University, Box 118,	SE-221 00 Lund, Sweden}


\begin{abstract}
We introduce a simple method for nearly simultaneous computation of all moments needed for quasi maximum likelihood estimation of parameters in discretely observed stochastic differential equations commonly seen in finance. The method proposed in this papers is not restricted to any particular dynamics of the differential equation and is virtually insensitive to the sampling interval. 
The key contribution of the paper is that computational complexity is sublinear in the number of observations as we compute all moments through a single operation. Furthermore, that operation can be done offline. 
The simulations show that the method is unbiased for all practical purposes for any sampling design, including random sampling, and that the computational cost is comparable (actually faster for moderate and large data sets) to the simple,  often severely biased, Euler-Maruyama approximation.
\end{abstract}


\end{frontmatter}


\section{Introduction}

Most applications such as simulation or estimation involving \Ito stochastic differential equations (SDEs) are in one way or another linked to the transition probabilities of the process. For example, it would be straightforward to estimate parameters using the maximum likelihood method if transition probability density was known, but this is rarely the case in practice. 

However, it is often possible to approximate the transition probability density. The probability density was obtained by brute force numerical computation of the solution to the Fokker-Planck equation (a partial differential equation) in  \cite{lo1988maximum,Erik07} while Monte Carlo based approaches were proposed in \cite{pedersen1995new,DurhamGallant,beskos2009monte,lindstrom2012regularized} and references therein. Those methods are computational expensive, making them unsuitable for large data sets. A Gauss-Hermite series expansion of the transition probability density was proposed by \cite{Ait}, although that approach is limited to models with a specific structure. 

The recent advances in collecting and storing large amounts of data are shifting the focus away from computationally slow but statistically efficient maximum likelihood methods towards computationally faster, yet not quite as statistically efficient quasi-maximum likelihood methods as the abundance of data often more than makes up for the loss of efficiency. 

A simple approach based on the quasi maximum likelihood technique was introduced in \cite{FlorensZmirou} where the  conditional mean and variance were obtained from an Euler-Maruyama discretization of the model, cf.\ \cite{kloeden1992numerical}. This is very efficient from a computational point of view and it was shown in \cite{FlorensZmirou} that their method is equivalent to the maximum likelihood estimator
as the sampling interval goes to zero, as the bias vanishes. A higher order version of this approach is proposed in \cite{Kessler97}.  Quasi maximum likelihood methods are generally unbiased, see \cite{sorensen2012estimating} provided that the mean and variance are correctly specified. The bias in the \cite{FlorensZmirou,Kessler97} methods therefore explicitly depends on the quality of the approximation of conditional moments, cf.\ \cite{Bachelier}.

The purpose of this paper is to develop a computationally fast method quasi maximum likelihood estimator for discretely observed diffusion processes that is suitable for moderate to large data sets. We show that the computational cost is sublinear rather than superlinear due to way the moments are computed. (our simulations shows the computational complexity is comparable to that of the Euler-Maruyama scheme, and hence magnitudes faster than any approximate maximum likelihood method). This will be achieved without the bias problems associated with the Euler-Maruyama method, a property that is virtually independent of the sampling interval. 

The outline of the paper is as follows. In Section \ref{sec:introduction} we formulate the statistical problem and discuss some alternative techniques for calculating conditional moments. This is followed by Section \ref{sec:DiscrKolmogorov} where we present a numerical implementation that results in sublinear complexity. The resulting parameter estimation algorithm  is demonstrated in Section~\ref{sec:parameterestimation} on two qualitatively different diffusion processes as well a randomly sampled data followed by the conclusions being drawn in Section \ref{sec:conclusions}.

\section{Diffusion processes and conditional moments}\label{sec:introduction}
Let $(\Omega,\mathcal{F}, P, \{ \mathcal{F}_t \}_{t \geq 0} )$ be a filtered probability space and let $X_t(\theta)$ be a stochastic process defined on that space that solves the following one dimensional stochastic differential equation (SDE)
\begin{equation}
dX_t = a_{\theta}(X_t) dt + b_{\theta}( X_t)dW_t , \quad X_{t_0} = x. \label{eq:SDE}
\end{equation}
We assume throughout the paper that the drift and diffusion terms are regular enough (e.g. bounded growth and local Lipschitz, see \cite{karatzas2012brownian} for alternative conditions) to ensure existence and uniqueness of the solution.

The optimal method for estimating the parameters, $\theta$, is the  maximum likelihood estimator. Let $x_k = x(t_k),~k=1, \ldots, K$ be observations generated from Eq~\eqref{eq:SDE}. The maximum likelihood estimator is defined as \begin{equation}
\hat{\theta}_{MLE} =\argmax_{\theta \in \Theta} \ell(\theta) \label{eq:MLEdef}
\end{equation} where the log-likelihood function is given by \begin{align}
\ell(\theta) & =   \log p_{\theta} (x_0) + \sum_{k=1}^K \log p_{\theta}(x_k| x_{k-1}).
\end{align} 

The transition probability densities, $p_{\theta}(x_k| x_{k-1})$ are implicitly defined by the model. The properties of the model are found by analysing the generator 
\begin{equation}
\mathcal{L} =   a_{\theta}(x) \frac{\partial}{\partial x}  + \frac{1}{2} b_{\theta}^2 (x) \frac{\partial^2}{ \partial x^2}. \label{eq:GeneratorSDE}
\end{equation} 

The transition probability $p_{\theta}(x_k| x_{k-1})$ is the solution to the Fokker-Planck equation, which is defined from the adjoint operator $(\mathcal{L}u, v) = (u,\mathcal{L}^*v)$ under the inner product $(\cdot,\cdot)$. The Fokker-Planck equation, when starting from $x_k$ at time $t_k$ and ending at $t_{k+1}$ is given by \begin{equation}
\frac{\partial }{\partial t}p_{\theta}(x,t) = -\mathcal{L}^* p_{\theta}(x,t) \label{eq:FP}
\end{equation} with the initial condition $p_{\theta}(x|x_{k}) = \delta(x - x_{k})$. This initial condition is likely to cause problems for numerical implementations of Eq.~\eqref{eq:FP} due to discontinuity, see the implementation in \cite{lo1988maximum} and the remedy proposed in \cite{Erik07}. 

Another method for computing the transition probability is to use the Markov property and law of total probability, adding and integrating out an intermediate state variable, see \cite{pedersen1995new,pedersen1995consistency} Let $t_{k-1} < s < t_{k}$. It then holds that \begin{align}
p_{\theta}(x_k| x_{k-1}) &= \int p_{\theta}(x_k, x_{s}| x_{k-1}) d x_{s}  \nonumber \\
	&= \mathbf{E}_{\theta}\left[ p_{\theta}(x_k | x_{s}) |  x_{k-1} \right]\label{eq:MonteCarloPDF}
 \end{align}	
 	Monte Carlo methods can easily approximate that expected value, but the use of variance reduction techniques is needed for most applications, cf.\ \cite{DurhamGallant,lindstrom2012monte}.
 	
 However, we cannot expect to be able to solve either the Fokker-Planck equation Eq.~\eqref{eq:FP} or the conditional expectation in Eq.~\eqref{eq:MonteCarloPDF} in closed form for more complex models. That means that the complexity of any of these approximate maximum likelihood method will be linear (in terms of expensive operations) in the number of observations. 
 	
 A possible remedy are the Gauss-Hermite, see \cite{Ait}, or saddle point, see \cite{ai2006saddlepoint} expansions. These can be very accurate for frequently sampled data but there are also important limitations. Such as  the existence of the Lamperti transform, and as well as $\Delta_k=t_k-t_{k-1}$ being small as the error typically is $\mathcal{O}\left( \Delta_k^{L+1} \right)$ with $L$ being the number of terms in the series expansion. A key operation is to employ a Lamperti transformation of the process \begin{equation}
 Y_t = g(X_t)
 \end{equation} such that the dynamics of $Z_t$ is given by \begin{equation}
 d Y_t = f(Y_t) dt + dW_t.
 \end{equation}
 The transition probability in \cite{Ait} is given by the Hermite series approximation (here assuming that $\Delta_k = \Delta$ for all observations)\begin{equation}
 p_Y(y_k|y_{k-1}) = \Delta^{1/2} \phi\left(\frac{y_k-y_{k-1}}{\Delta^{1/2}}\right) \sum_{j=0}^{\infty} \eta^{\theta}_j H_j\left(\frac{y_k-y_{k-1}}{\Delta^{1/2}}\right)
 \end{equation} with the coefficients given by \begin{equation}
\eta^{\theta}_j = \frac{1}{j!} \ee_{\theta} \left[H_j\left(\frac{y_k-y_{k-1}}{\Delta^{1/2}}\right) | z_{n-1} \right]
 \end{equation} where $H_j$ is a Hermite polynomial of order $j$. It is worth noting that the series expansion will not converge for all distributions and that a finite expansion may be negative for some values. The latter can be solved by considering a series expansion of the log-density, but that series may not integrate to unity. Still, the complexity is essentially sublinear as the major complexity, deriving the expansion, is performed only once.
 	
 A simpler alternative is to resort to a quasi maximum likelihood estimator, cf.\ \cite{godambe2010quasi,sorensen2012estimating,lindstrom2015statistics}. The downside is a loss of statistical efficiency as the distribution of the Maximum likelihood estimate is given by \begin{equation}
 \sqrt{N} \left(\hat{\theta} - \theta_0 \right) \stackrel{d}{\rightarrow} N(0, I_F^{-1}),
 \end{equation} where $\theta_0$ is the true parameter and  $I_F$ is the Fisher information matrix defined as \begin{equation}
  (I_F)_{i,j} = \ee \left[  \frac{\partial^2}{\partial \theta_i \partial \theta_j}  \log p(X|\theta_0) \right]
 \end{equation} or equivalently \begin{equation}
(I_F)_{i,j} =  \ee \left[  (\frac{\partial}{\partial \theta_i} \log p(X|\theta_0))  (\frac{\partial}{\partial \theta_j}  \log p(X|\theta_0))  \right]
 \end{equation} whereas the distribution of the quasi maximum likelihood estimate is given by \begin{equation}
\sqrt{N} \left(\hat{\theta} - \theta_0 \right) \stackrel{d}{\rightarrow} N(0, J^{-1} I J^{-1}),
\end{equation} where $$(J)_{i,j} = \ee \left[  \frac{\partial^2}{\partial \theta_i \partial \theta_j}  \log \Psi(X|\theta_0) \right]$$ and $$(I)_{i,j} = \ee \left[  ( (\frac{\partial}{\partial \theta_i}  \log \Psi(X|\theta_0))  (\frac{\partial}{\partial \theta_j}   \log \Psi(X|\theta_0))  \right]$$ with $\Psi$ being a Gaussian density with location and scale parameters given by the conditional mean and (co)variance. That covariance is always larger or equal to the variance of the maximum likelihood estimate (this follows from the Cramer-Rao inequality), with the difference typically being rather small for nearly Gaussian models, cf. \cite{overbeck1997estimation}.
 
\subsection{Conditional moments}
Parameter estimation using QML is a reason for bringing us to the topic of calculating conditional moments of the stochastic process.
We will throughout this section assume that $g(\cdot) $ is a general function representing any conditional moment of interest. The conditional moment is given by
\begin{equation}
\mathbf{E}_{\theta}[g(X_k)| X_{k-1} = x_k]  = \int g(x_k) p_{\theta}(x_k| x_{k-1}) d x_k,\label{eq:GeneralCondExp}
\end{equation}
where $ p_{\theta}(x_k| x_{k-1})$ is the conditional probability density.

The discussion in the previous section illustrates why Eq.~\eqref{eq:GeneralCondExp} is intractable in the general case, which is why we have to resort to approximations.  Most approximations of a conditional moment can be expressed as a weighted sum
\begin{equation}
\mathbf{E}_{\theta}[g(X_{k})| X_{k-1} = x_k] \approx \sum_i \omega_i g(\xi_i).
\end{equation}
This approximation includes techniques like Monte Carlo estimation and various deterministic quadrature rules, such as rectangular rule, the trapezoidal rule of the Gauss-Hermite quadrature. 

It is also possible to approximate that conditional expectation using an \Ito-Taylor expansion. Assume that  the function $g$ is $2 k+1$ times continuously differentiable. It then holds that
\begin{equation}
\mathbf{E}_{\theta}[g(X_k)| X_{k-1} = x_k] = \sum_{l=0}^L \mathcal{L}^l g(x_k) \frac{\Delta_k^l}{l!} + \mathcal{O}(\Delta_k^{L+1}),\;\; \Delta_k = t_k-t_{k-1} \label{eq:gen}.
\end{equation}
Note that this expansion is not guaranteed to converge unless additional constraints are imposed on $X$, see \cite{Ait} for details, but it often works quite well for small time intervals as 
the leading error term is $\mathcal{O}(\Delta_k^{L+1})$. This type of approximation is used compute moments in the Hermite series expansion in \cite{Ait}. It may be necessary to iterate the \Ito-Taylor expansion over a series of smaller steps $\Delta t/m$ for sparsely sampled data, cf. Runge-Kutta and multistep methods, see \cite{kloeden1992numerical}. 

An alternative, mentioned in the beginning, is to calculate conditional moments using the generator. This will require us to solve one PDEs  for each moment. The Feynman-Kac (F-K) formula, see \cite{karatzas2012brownian}, establishes the relation between conditional expectations and parabolic partial differential equations. Specifically, let $\tau \in [t_{k-1}, t_k]$ and define the conditional expectation \begin{equation}
u(x,\tau) = \mathbf{E}_{\theta}[g(X_k)| X_{\tau} = x]
\end{equation} when the dynamics is given by Eq.~(\ref{eq:SDE}). The solution to the expectation is then given as the solution to \begin{equation}
\frac{\partial }{\partial \tau}u(x,\tau) = -\mathcal{L} u(x,\tau). \label{eq:FK}
\end{equation} where the operator $\mathcal{L}$ is defined in Eq.~(\ref{eq:GeneratorSDE}) and the initial condition is given by $u(x,t_{k}) = g(x)$.

 There are at least two advantages of solving the adjoint problem compared to the Fokker-Planck equation. The first is that we only need to solve one single PDE for any number of observations, which should be compared to the computational complexity of the Monte Carlo and quadrature methods where it is necessary to propagate weights and/or particles between each observation. Secondly, it is more robust from a numerical point of view to solve the adjoint equation as it has a well posed initial condition (the equation is solved backwards in time) e.g.\ for the standard moments a polynomial, $x^p$. 

We will later show that it is in fact enough to solve a single PDE regardless of the number of moments we are interested in. That makes the computational complexity  marginal compared to that of a full blown approximate maximum likelihood estimator.

We present a conceptual summary of the pros and cons of each method respectively in Table~\ref{tab:comparemethods}. We have marked the Fokker-Planck method with a $(*)$  since the performance of this method for small $\Delta_k$ depends strongly in the type of initial condition as described earlier.

\begin{table}[h]
	\caption{Feasibility of different methods for parameter estimation. }
	\label{tab:comparemethods}
	\begin{center}
		\begin{tabular}{lllll}
			\multicolumn{1}{c}{\bf Method}  &\multicolumn{1}{c}{\bf Small $\Delta$ }&\multicolumn{1}{c}{\bf Large $\Delta$ }  &\multicolumn{1}{c}{\bf Small data set} &\multicolumn{1}{c}{\bf Large data set}
			\\ \hline \\
			Euler-Maruyama        & Yes  & - & Yes & Yes \\
			\Ito-Taylor  (Eq.~\eqref{eq:gen})               & Yes  & - & Yes & Yes \\
			Fokker-Planck      & $*$ & Yes & Yes & - \\
						Monte Carlo      & Yes & Yes & Yes & - \\
						Hermite series & Yes & - & Yes & Yes \\
			Generator (F-K)      & Yes  & Yes & Yes & Yes \\
		\end{tabular}
	\end{center}
\end{table}

In the following section we will present a numerical 
approach to calculate the conditional moments from the Kolmogorov-backward equation.



\section{Discretization of the Kolmogorov backward equation}\label{sec:DiscrKolmogorov}
\label{sec3}
Solving the Kolmogorov backward equation numerically can be achieved with a large number of different methods. 
We have opted for a semi-discretization with central differences in space to achieve  maximum simplicity and as we will later see also the possibility to reuse calculation.
The backward equation  Eq.~(\ref{eq:FK}) is a Cauchy problem in the sense that it has a final condition  defined by the conditional moment of interest, $u(x,T) = g(x)$, but lacks boundary conditions. This is similar to many option pricing problems where it is common to impose a
boundary condition from asymptotic expansion of the solution, cf. \cite{BenchOp2015} for an overview of numerical techniques for computing option prices. For Eq.~(\ref{eq:FK}) to be well-posed it is necessary to impose boundary conditions for certain values of the coefficients. The condition when this is necessary may be found from the Fichera function (here in one dimension), 
\begin{equation}
\mathcal{F}ich = \sum_{i}^{0,N}\left(a_{\theta}(x_i)  + \frac{1}{2} \frac{\partial}{ \partial x}b_{\theta}^2 (x_i) \right)\kappa
\end{equation}
where $\kappa = \{-1,1\}$ are the boundary normals. Boundary conditions are not required when $\mathcal{F}ich\geq 0$, \cite{Fichera56}.
As an example the Fichera condition for the Cox-Ingersoll-Ross model at $x_0 = 0$ is given by $ab \geq \sigma^2/2$ which is also known as the Feller condition.
Under the assumption that the backward equation is well-posed without boundary conditions in the sense of positive Fichera, we still need to define some conditions for the boundary values for the semi-discretized system. In \cite{BC} it was suggested to calculate the boundary values by solving a simplified 
backward equation at the boundaries with finite differences defined on the internal node points. We generalize this approach here by solving 
 the full equation Eq.~(\ref{eq:FK})  at the boundary using interior node points. The advantage of this approach is that it does not require a large solution domain and it does not introduce a right hand side vector in the algebraic system. This enables rapid calculation of the time integration of the solution which we will utilize.
 
Turning to the discretization of the derivatives for the interior nodes.  The first and second partial derivatives are approximated by second order central differences with $u_{n} = u(x_{\min} + n h)$, $h$ being the distance between two nodes. The derivatives are then given by
\begin{equation}
\frac{\partial u_n}{\partial x} \approx \frac{1}{2h}\left( u_{n+1} - u_{n-1} \right)\label{eq:FDb1}
\end{equation}
and
\begin{equation}
\frac{\partial^2 u_n}{\partial x^2} \approx  \frac{1}{h^2}\left( u_{n+1} - 2u_{n}  + u_{n-1}  \right).\label{eq:FDb2}
\end{equation} with both approximations having errors of size $\mathcal{O}(h^2)$.
Inserting the FD approximations (\ref{eq:FDb1} and \ref{eq:FDb2}) into Eq.~(\ref{eq:FK}) results in
\begin{equation}
 \frac{\partial u_n}{\partial \tau} = -a_{\theta}(x_n) \frac{1}{2h}\left( u_{n+1} - u_{n-1}  \right)  - \frac{1}{2h^2}b_{\theta}^2(x_n) \left( u_{n+1} - 2u_{n}  + u_{n-1}  \right)\label{eq:FD}.
\end{equation}
At the boundaries $0,N$ we need to solve Eq.~(\ref{eq:FK}) with skewed finite differences. On the lower boundary we use the following scheme:
\begin{equation}
\frac{\partial u_n}{\partial x} \approx -\frac{1}{h}\left( \frac{3}{2}u_0 - 2 u_1 +\frac{1}{2}u_2 \right) \label{eq:FDleft1}
\end{equation}
and
\begin{equation}
\frac{\partial^2 u_n}{\partial x^2} \approx \frac{1}{h^2}\left( 2u_0 - 5 u_1 +4u_2 - u_3\right).\label{eq:FDleft2}
\end{equation}
Similar approximations are used on the upper boundary, 
\begin{equation}
\frac{\partial u_n}{\partial x} \approx \frac{1}{h}\left( \frac{3}{2}u_N - 2 u_{N-1}+\frac{1}{2}u_{N-2} \right) \label{eq:FDright1}
\end{equation}
and
\begin{equation}
\frac{\partial^2 u_n}{\partial x^2} \approx \frac{1}{h^2}\left( 2u_N - 5 u_{N-1} +4u_{N-2} - u_{N-3}\right) .\label{eq:FDright2}
\end{equation}
Inserting these FD approximations (\ref{eq:FDleft1}-\ref{eq:FDright2}) into Eq.~\eqref{eq:FK} result in similarly equations as Eq.~\eqref{eq:FD}.
Our approximation to the Kolmogorov-backward equation after approximating the spatial operator is given by, 
\begin{equation}
\frac{\partial u_n}{\partial \tau} = \mathbf{A} u_n(\tau) \label{eq:discretized}
\end{equation}
where  $\mathbf{A}$ is a banded matrix with the following elements,
\begin{eqnarray*}
&A_{i,i+1} =  \frac{a_{\theta}(x_i)}{2h} - \frac{b_{\theta}(x_i)^2}{ 2h^2}, \; A_{i,i}  =  \frac{b_{\theta}(x_i)^2}{h^2}, \;A_{i,i-1} =  -\frac{a_{\theta}(x_i)}{2h} -\frac{b_{\theta}(x_i)^2}{2 h^2}.
\end{eqnarray*}
The first and last rows in $\mathbf{A}$ have extra nonzero columns from the extrapolated boundary equations.
Now that we have discretized the spatial operator we turn to the time discretization. In this paper we will use the matrix exponential to propagate in time, see \cite{expmap}. This is feasible due to the boundary technique introduced earlier. To illustrate the benefit of our approach of including the boundary values in $\mathbf{A}$  we consider the case when standard boundary conditions e.g. Dirichlet are used. The semi-discretized system then become
\begin{equation}
\frac{\partial u_n}{\partial \tau} = \mathbf{\tilde{A}} u_n(\tau) + \mathbf{b} \label{eq:discretized2}
\end{equation}
where $\mathbf{b}$ contains the boundary values (here assumed to be time independent). The general solution of Eq.~\eqref{eq:discretized2} is given by 
\begin{equation}
u_n(t_{k-1})  = \exp\left( \mathbf{\tilde{A}} (t_{k-1}-t_k)\right) u_n(t_k) + \mathbf{\tilde{A}}^{-1}\left( \exp\left( \mathbf{\tilde{A}} (t_{k-1}-t_k)\right) -\mathbf{I}  \right)\mathbf{b} \label{eq:gensol2}
\end{equation} 
which is computationally more expensive then the solution of Eq.~\eqref{eq:discretized},
\begin{equation}
u_n(t_{k-1}) = \exp\left( \mathbf{A} (t_{k-1}-t_k)\right) u_n(t_k)  \label{eq:gen1}.
\end{equation}
Furthermore another drawback with the classical boundary values is $\dim(\mathbf{\tilde{A}}) \gg \dim(\mathbf{A})$ since it requires a larger solution domain to avoid boundary values influencing the solution which is an additional computational cost.
Returning to  Eq.~\eqref{eq:gen1} the approximation error to the analytical solution (in terms of an exponential map) is given by,
\begin{align}
u(x, t_{k-1}) &= \exp\left(\mathcal{L} (t_{k-1}-t_k)\right) u(x, t_k)  \\
&\approx  \exp\left( \mathbf{A} (t_{k-1}-t_k)\right) u_n(t_k) = u_n(t_{k-1})  + \mathcal{O}(h^2).   
\end{align} 
Since $\mathbf{A} $ is not normal we might need to subiterate the solution in time for stability reasons.  This is also required
when we need to evaluate the solution at non-equidistant time intervals e.g. when the data is collected at random time instances. A subiterated solution is obtained from the following identity $\exp\left( \mathbf{A}\tau\right)   = \left( \exp\left( \mathbf{A} \tau/m\right) \right)^m$ where a typical good value for $m$ can be found from the following condition $\| \mathbf{A} \tau/m \| \leq 1 $ given in \cite{expmap}.

We use cubic spline interpolation to compute the conditional expectation for values that are not part of the finite difference grid. The interpolation error due to the cubic splines are $\mathcal{O}(h^4)$ which is dominated by the finite difference error for a dense grid.

The conditional mean and variance are accurately computed from the solution of the first moment $$\hat{u}^{(1)}(x, t_{k-1} )  = \hat{\ee}[X_k| X_{k-1} = x]  $$ and second moment
 $$\hat{u}^{(2)}(x, t_{k-1} )  =  \hat{\ee}[X^2_k| X_{k-1} = x]. $$

The conditional variance is then obtain through a combination of these
\begin{align}
\widehat{\var}[ X_k |  X_{k-1} = x] & = \hat{\ee}[X^2_k |  X_{k-1} = x] -  \hat{\ee}^2[X_k |  X_{k-1} = x] \nonumber \\
& = \hat{u}^{(2)}(x, t_{k-1} ) - \hat{u}^{(1)}(x, t_{k-1} )^2.
\end{align}

\subsection{Convergence}
The semi discretization does not introduce any errors due to the time integration. However, the discretization of the derivatives and the interpolation do introduce errors. We can decompose the interpolated numerical solution $\hat{u}^{interp}$ by adding and subtracting the numerical solution without interpolation $\hat{u}$ and the true solution $u$. This leads to\begin{align}
\hat{u}^{interp} &= \hat{u}^{interp} \pm \hat{u} \pm u \nonumber  \\
&= \underbrace{\hat{u}^{interp} - \hat{u}}_{Interpolation,~\mathcal{O}(h^4)} + \underbrace{\hat{u} - u}_{Discretization,~\mathcal{O}(h^2)} + u \nonumber  \\ 
&= u + \mathcal{O}(h^2).
\end{align} Hence, the interpolation error is dominated by the discretization error if a good interpolation method is used. That trivially implies that the conditional mean can be computed with arbitrary accuracy. 

Next, we find that the conditional variance is given by \begin{align}
\hat{u}^{(2),interp} - (\hat{u}^{(1),interp})^2 & = \left(u^{(2)} + \mathcal{O}(h^2) \right) - \left( u^{(1)} + \mathcal{O}(h^2) \right)^2 \\
&= u^{(2)} -  (u^{(1)})^2 + \mathcal{O}(h^2) 
\end{align} meaning that also the error in the conditional variance is controlled by the denseness of the finite difference grid, $h$. This means that both the error in the mean and covariance can be made arbitrarily small (we choose the design parameter $h$), leading to consistent estimates, cf. \cite{sorensen2012estimating}. This is in contrast to the approximate QML estimators in \cite{FlorensZmirou} and \cite{Kessler97} where no refinement of the estimates are possible.

The numerical quality of the method is benchmarked by comparing it against the conditional mean and variance of the Cox-Ingersoll-Ross (CIR) and the conditional mean of its inverse (iCIR). These models are defined by,
\begin{align}
dX_t &= a(b - X_t)dt + \sigma X_t^{1/2} dW_t\label{eq:CIR} & \textbf{CIR}\\
d\tilde{X}_t& = \left[a\tilde{X}_t + (\sigma^2 -ab )\tilde{X}_t^2 \right]dt -\sigma \tilde{X}_t^{3/2} dW_t\label{eq:iCIR} & \textbf{iCIR}
\end{align} with $\tilde{X}_t = 1/ X_t$. That means that we can (at least conceptually) compute the transition probability density as well as moments for the iCIR process.

\begin{figure}
\centering
  \begin{subfigure}[b]{.50\linewidth}
    \centering
    \includegraphics[width=0.99\textwidth]{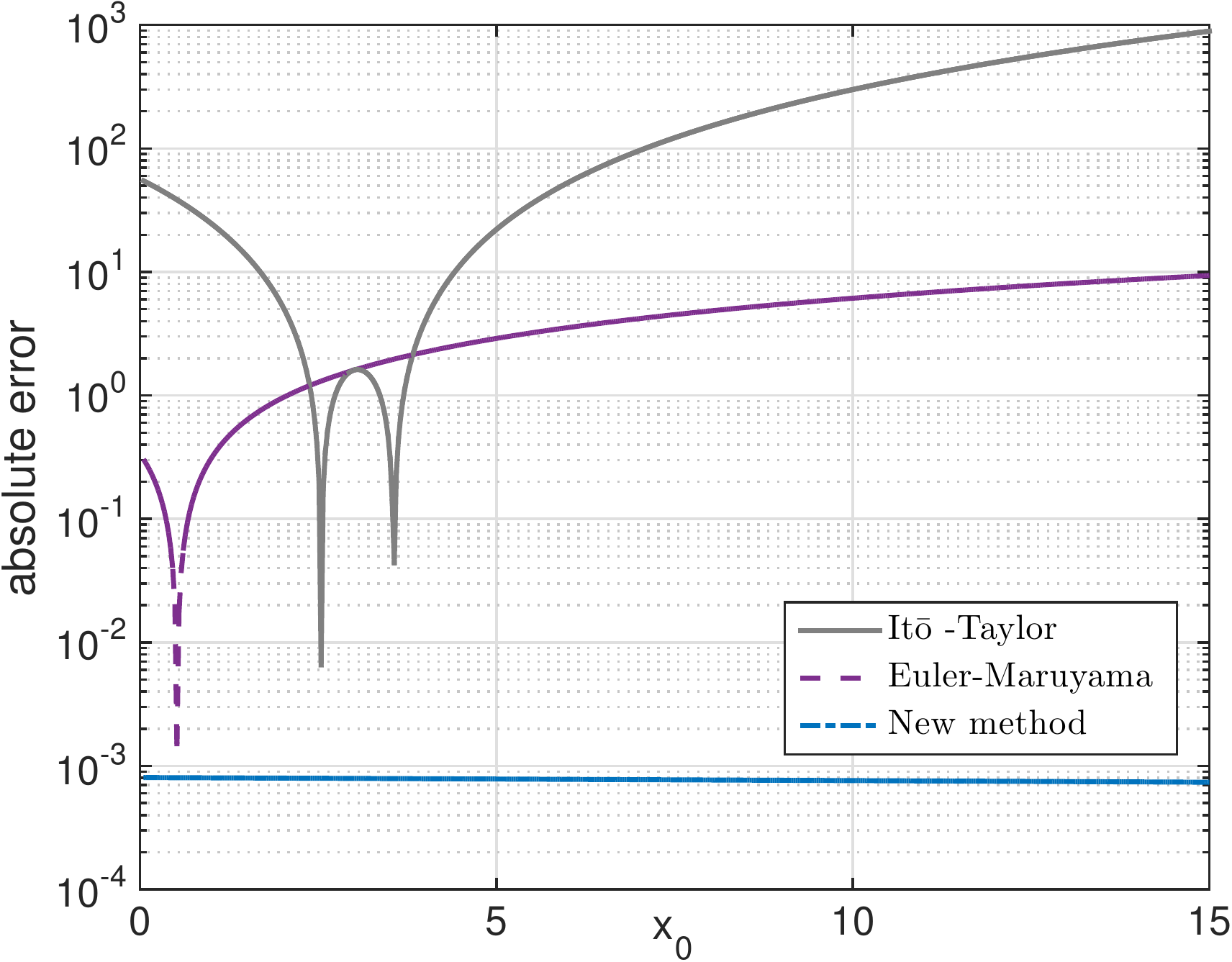}
	\caption{Error in the conditional mean}   \label{fig:ErrorMean}
  \end{subfigure}%
  \begin{subfigure}[b]{.50\linewidth}
    \centering
    \includegraphics[width=0.99\textwidth]{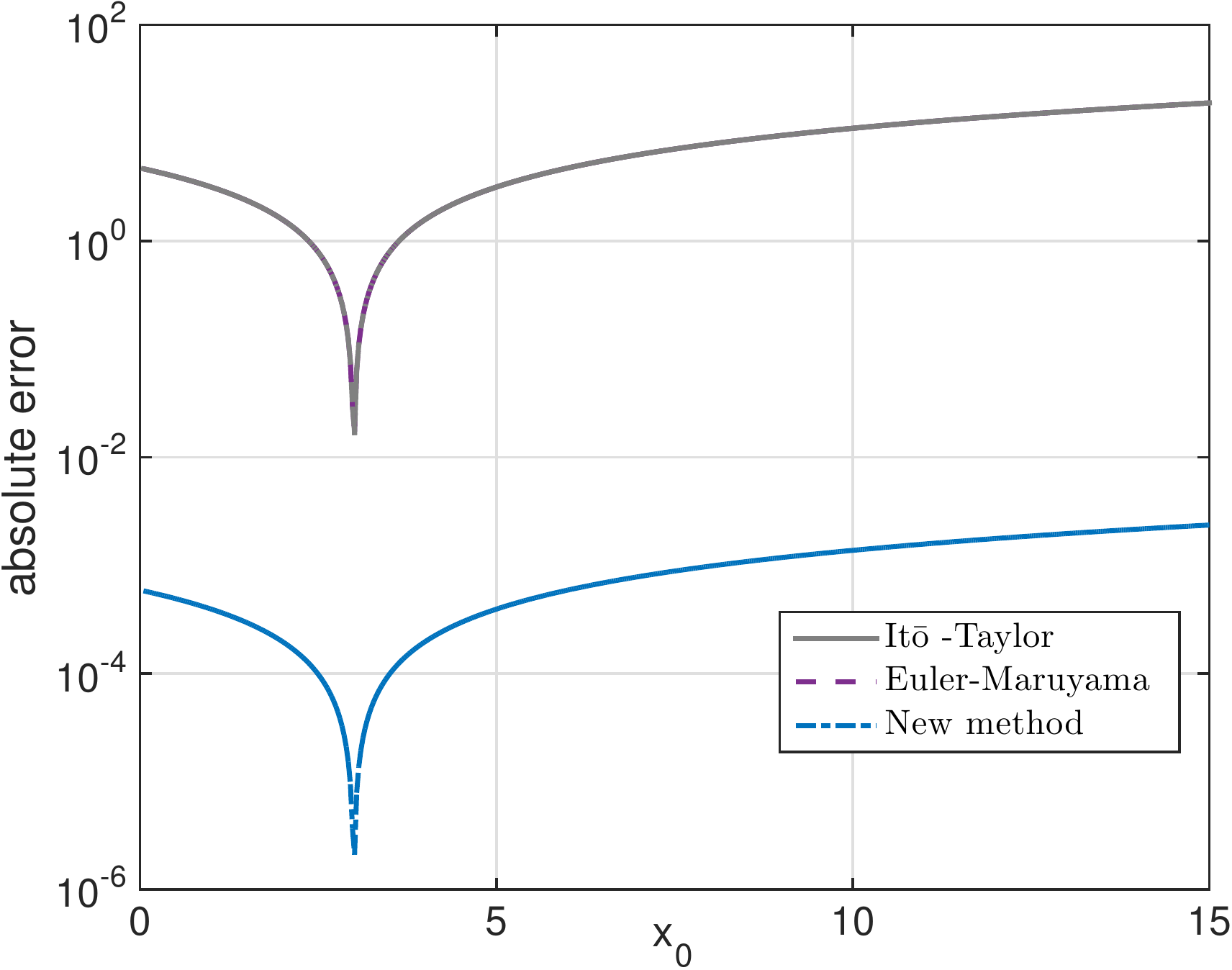}
  \caption{Error in the conditional variance}\label{fig:ErrorVar}
  \end{subfigure}
\caption{Absolute error for the CIR model compared between different methods for $T= 1/6$. The methods are the Euler-Maruyama scheme, truncated  \Ito-Taylor ($k=1$) Eq.~(\ref{eq:gen}) and the proposed method. Note that the conditional var (right figure) is equal for the Euler-Maruyama and the \Ito-Taylor while the mean (left figure) is not. The time $T$ was selected such that all methods converged. The Euler-Maruyama and the generator approximation would perform even worse if larger $T$ is used.}\label{fig:ICIRabserror}
\end{figure}

Let $x_{n} \in \{  x_{\mathrm{min}}, \ldots, x_{\mathrm{max}} \}$ be a grid with $x_{\mathrm{min}}  = 0.05$ and $x_{\mathrm{max}} = 0.15$ and $t_m \in \{0, \ldots, T\}$ where the final time $T = 1/6$. The conditional mean of the iCIR process is quite lengthy and involves a Gamma function and will not be expressed here, see \cite{ICIR}. The absolute error between the proposed method, conditional moments using the \Ito-Taylor expansion and the Euler-Maruyama scheme are compared to the exact moments for the CIR model in Figure~\ref{fig:ICIRabserror}. 

\begin{figure}
\centering
    \includegraphics[width=0.7\textwidth]{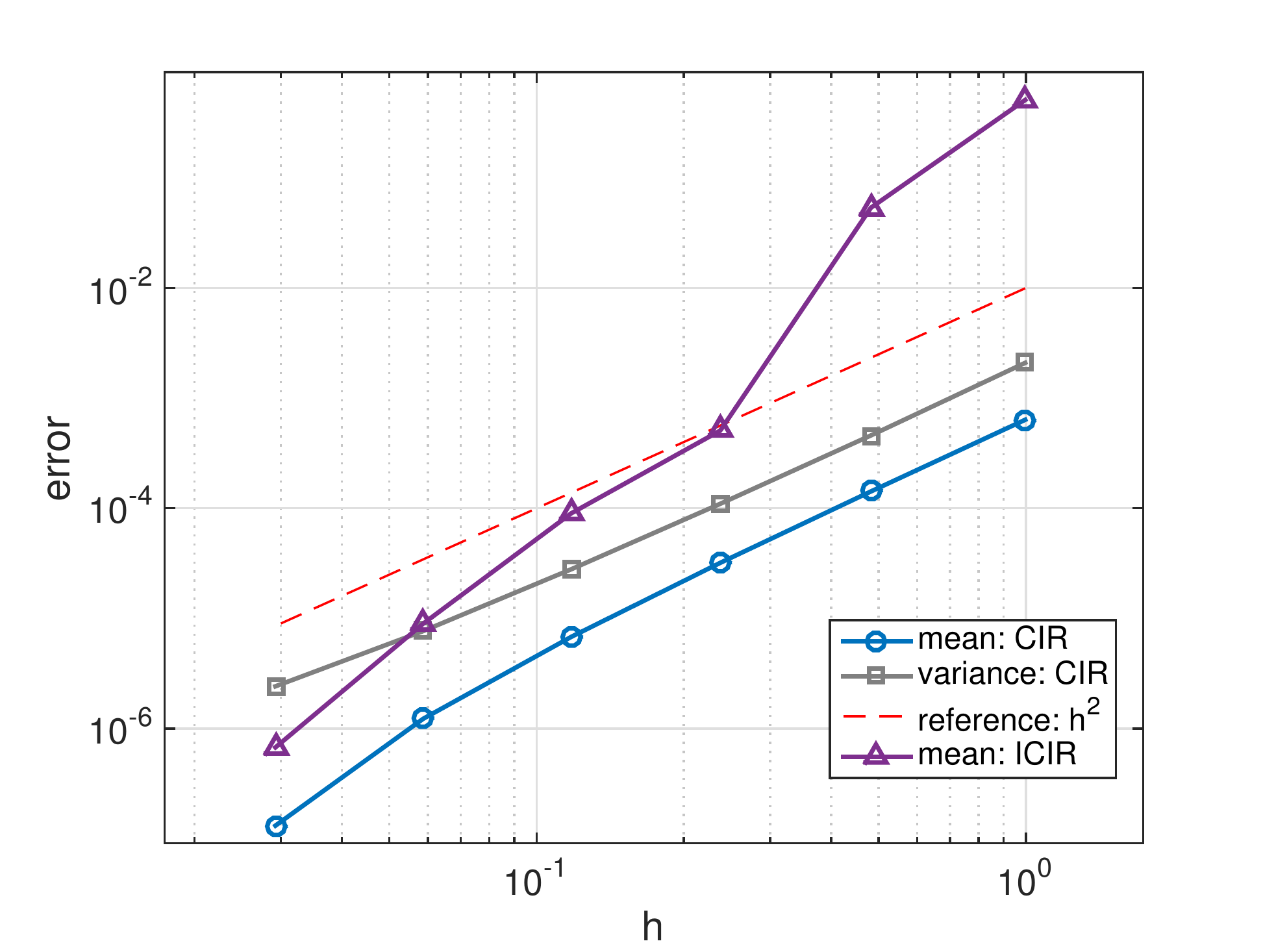} 
\caption{Spatial relative mean square error of the conditional mean for the iCIR and CIR model and conditional variance for the CIR model. Here $\tau= 1/6 $ and number of time steps are $N_t= 1000$. Parameters for this test 
was $\{a,b,\sigma\} = \{15,3, 2\}$.}  \label{fig:CIRabserrors}
\end{figure}

The convergence of the numerical method is seen in Figure~\ref{fig:CIRabserrors} where we have plotted the relative error (relative mean square error) as a function of spatial discretization for the CIR model using parameters $N_x = 2^{4:9}-1$  with $h = (x_{\mathrm{max}}- x_{\mathrm{min}})/N_x$.


\section{Parameter estimation}\label{sec:parameterestimation}

To test the applicability of the proposed framework we evaluate the quasi maximum likelihood estimator on two diffusion models. 
The quasi maximum likelihood estimator is defined as \begin{equation}
\hat{\theta} = \argmax_{\theta \in \Theta} \sum_{k=1}^K \log \Psi \left( x_k; \hat{\ee}_{\theta}[x_t|x_{k-1}], \widehat{\var}_{\theta}[x_k|x_{k-1}] \right)
\end{equation}
where $ \Psi \left( x_k; \hat{\ee}_{\theta}[x_k|x_{k-1}], \widehat{\var}_{\theta}[x_k|x_{k-1}] \right)$
is the Gaussian density function with mean $ \hat{\ee}_{\theta}[x_k|x_{k-1}]$ and variance $\widehat{\var}_{\theta}[x_k|x_{k-1}]$ computed with the new method.  We compare the moments computed from the adjoint equation in Section~\ref{sec3} with the Euler-Maruyama method and to the exact moments when they are known.

\subsection{Estimation on moderate data set}\label{sec:ModerateData}

We also consider the  inverse CIR model commonly used in interest rate modeling, see Eq.~(\ref{eq:iCIR}). This model (or actually a simplification of it) was the preferred model in for US interest rate data in the likelihood based analysis in \cite{durham2003likelihood}. The model is challenging as the drift is non-linear but it can be shown that the conditional moments can be calculated  analytically. The test data was generated from the inverse CIR model using monthly time steps $\Delta t = 1/12$. Using $x_0=5$ initial value, we generated $N=1000$ observations using the parameter $\{a, b, \sigma \} = \{15, 3, 2\}$ . The first $100$ observations were then discarded as burnin, leaving us with $900$ observations. This was repeated $100$ times in order to evaluate the estimators on independent data sets.

\begin{figure}[htbp]
\begin{center}
\includegraphics[width=\textwidth]{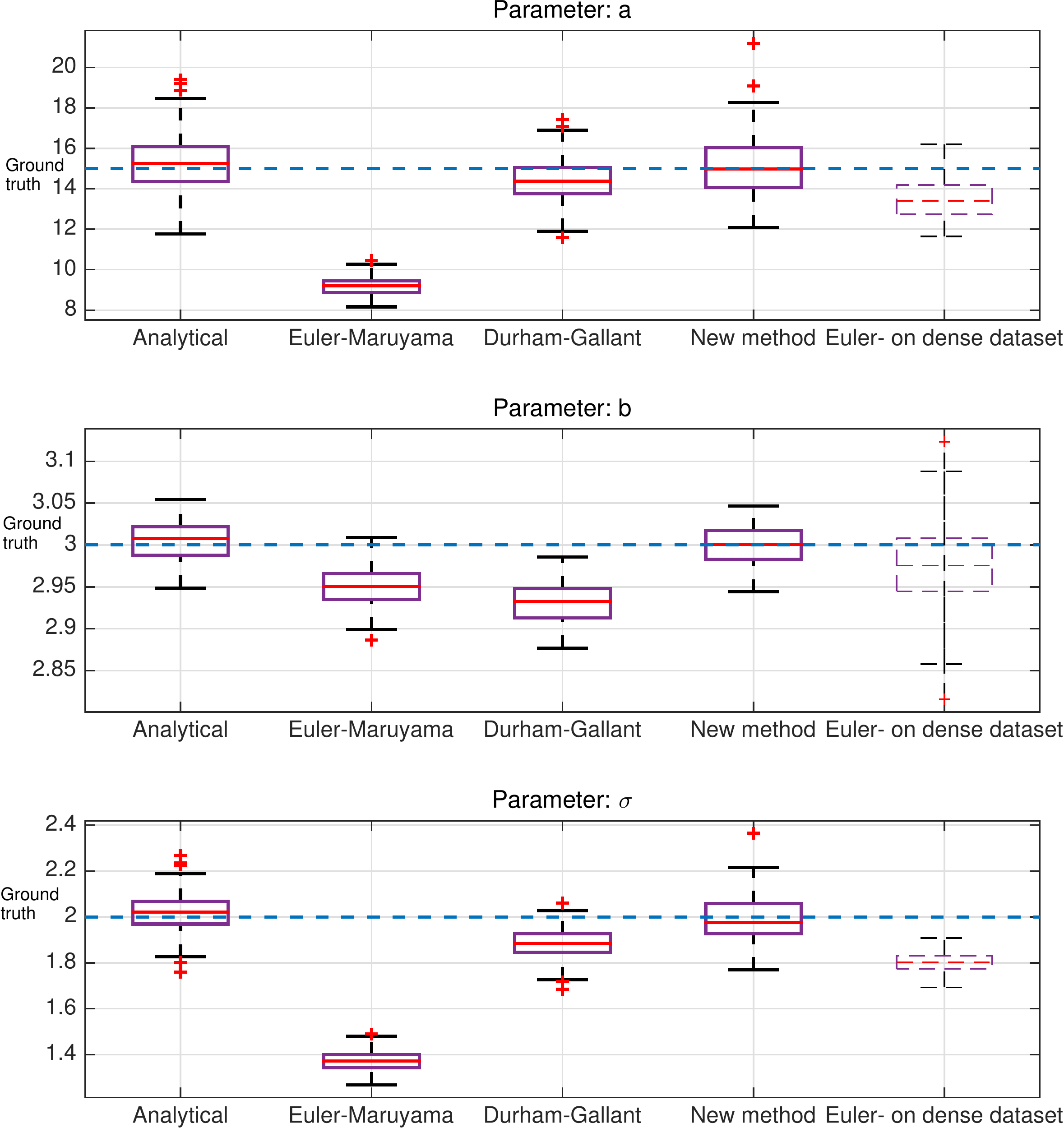}
\end{center}
\caption{Estimated parameters for the iCIR model using an analytical expression for the likelihood function, the Euler Maruyama approximation, the Durham-Gallant method, our proposed method and an Euler Maruyama estimator using more densely sampled data. The data sampling interval was $\Delta t = 1/12$ for the iCIR model. Since this time step resulted in a large bias in the Euler-Maruyama method; We also ran it on a data set with $\Delta t=1/52$, dashed boxes in the Figure. For this dense data set the Euler-Maruyama method performed better, but still suffers from bias. }\label{fig:iCIRstd}
\end{figure}

The estimation was conducted within the quasi maximum likelihood framework on an Intel\textsuperscript{\textregistered}  Core i5 @ 2.5 Ghz with 8GB of RAM.  As an optimizer we used  the standard  Nelder-Mead, (fminsearch) in Matlab\textsuperscript{\textregistered} (R2014b) with initial guess $\{10, 5, 1\}$. The results for the iCIR process is presented in Figure~\ref{fig:iCIRstd} where we see that the proposed method is unbiased whereas the Euler-Maruyama as well as the Durham-Gallant, see \cite{DurhamGallant}, approximate maximum likelihood estimator are biased (the latter is due to insufficient imputation in the time domain). We also see that the proposed method is as virtually as good as the maximum likelihood estimator based on the closed form transition probability density. We also note that the Euler-Maruyama is still worse than the proposed method even when a more densely sampled data set is available for that estimator.

\subsection{Estimation on randomly sampled data sets }

\begin{figure}[htbp]
\begin{center}
\includegraphics[width=\textwidth]{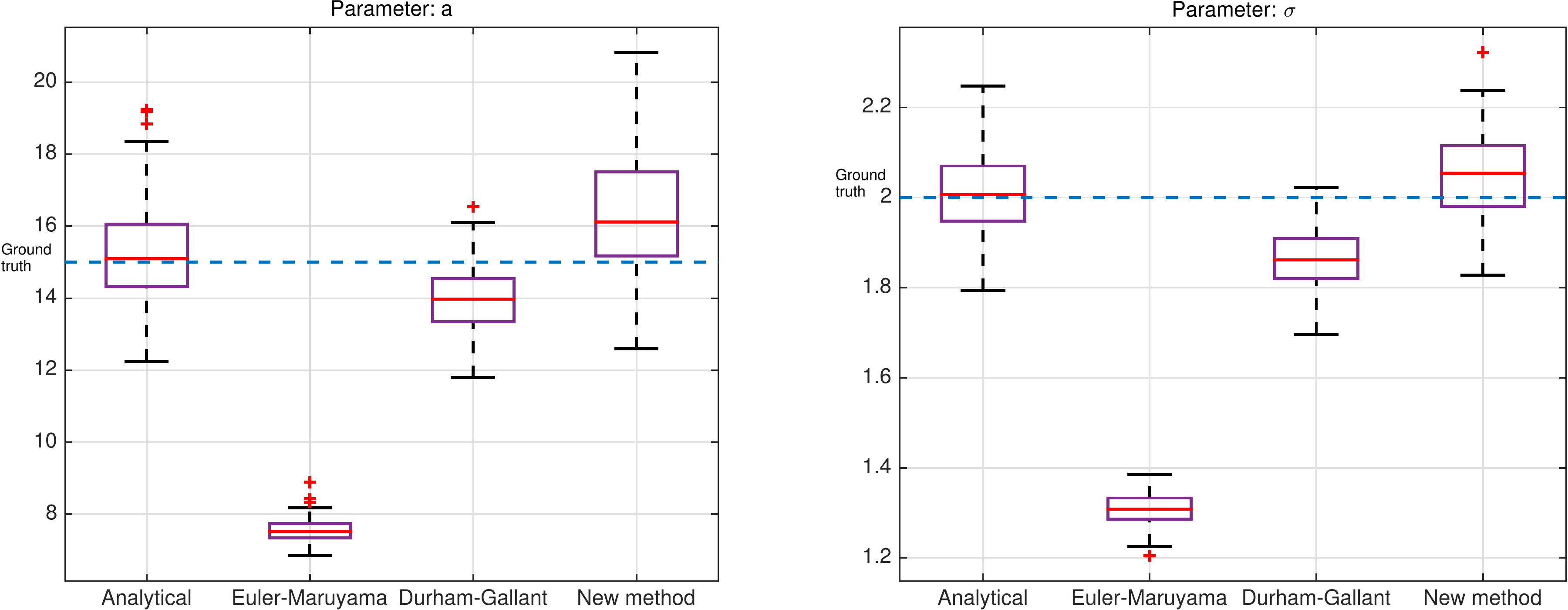}
\end{center} 
\caption{\label{fig:CIRrandom} Estimated parameters for the CIR model with randomly sampled data using an analytical expression for the likelihood function, the Euler Maruyama approximation, the Durham-Gallant method and our proposed method. The data sampling interval was randomly distributed as $\Delta t \sim U([1/252, 1/6])$.   }
\end{figure}

To further test the applicability of the proposed method we estimate parameters on a simulated data set from a CIR process with 
samples arriving at random times, cf. \cite{ait2003effects}. This would be a challenging problem for a discrete time model, but can readily be handled with a continuous time model within the proposed framework. We have simulated $1000$ observations from the CIR process, cf.\ Eq.~(\ref{eq:CIR}) with the same parameters and burnin as for the iCIR process with random time intervals. The time intervals are uniformly distributed $t_k-t_{k-1} \sim U([1/252,1/6])$. The results when estimating the parameters with the proposed method, Durham-Gallant method and the Euler-Maruyama is presented in Figure~\ref{fig:CIRrandom} (note that we only present the $a$ and $\sigma$ parameters as the $b$ essentially is given by the unconditional mean of the data). It can be seen that the proposed method is unbiased even for randomly sampled data, but also that the variance for the proposed method is worse than that of the maximum likelihood estimator based on the analytical transition probability density. This is in line with our intuition as the transition density will become less Gaussian for sparsely sampled data, making the MLE the preferred estimator in that situation. 

\subsection{Estimation on large data sets }

The challenge of large data sets is to develop fast methods. The Euler-Maruyama method is very fast (moments are given in closed form) but requires $\Delta t \rightarrow 0$ and $K \Delta t \rightarrow \infty$ for consistency, cf.\ \cite{sorensen2012estimating}. The proposed method computes practically unbiased estimates, cf.\ Figures \ref{fig:iCIRstd} and \ \ref{fig:CIRrandom}, for any sampling interval when the finite difference grid is dense enough, and will therefore only require $K \rightarrow \infty$ for consistency. 

\begin{figure}[htbp]
\begin{center}
\includegraphics[width=0.88\textwidth]{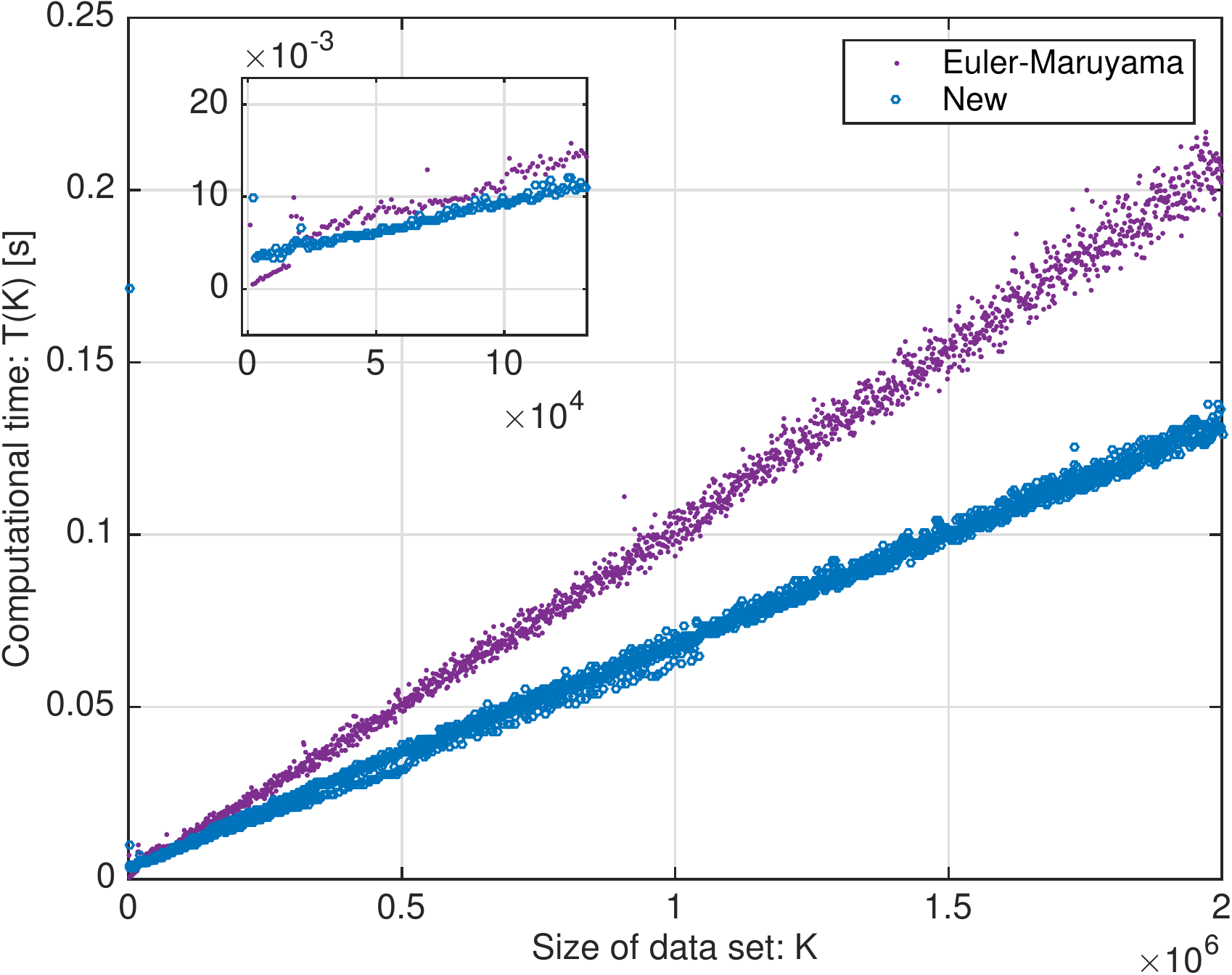}
\end{center}
\caption{Computational time (wall-clock time) averaged over three consecutive measurements as a function of the data set size.   }\label{fig:TimePlot}
\end{figure}

Here we evaluate the computational performance when computing the quasi likelihood function when the data consists of $K=2~000~000$ observations, which makes the data set computationally infeasible for most other estimators. The parameters and sampling is the same as in Section~\ref{sec:ModerateData}.
We present the time needed to compute the quasi likelihood function for an increasing set of observations (each point in the graph represents $1000$ additional observations) for the Euler-Maruyama and the proposed method in Figure~\ref{fig:TimePlot}. The plot is based on the average taken over three simulations.
The proposed method is initially more expensive than the Euler-Maruyama as we need to compute one matrix exponential to obtain the moments. However, the cost after the initial computation scales similarly as for Euler-Maruyama (it is actually somewhat cheaper for many observations), in spite that our method is consistent while the Euler-Maruyama is severely biased.
This result is very encouraging as we do not need to have frequently sampled data for consistency meaning that we can work with data sets sampled over longer time horizons at the same computational cost, meaning that we could estimate certain (typically drift parameters) much better than what would be possible using only the Euler-Maruyama or similar algorithms.

\section{Conclusions}\label{sec:conclusions}
This paper introduces a framework for computing conditional moments for diffusion models based on numerical computation of the Kolmogorov-backward equation which is the adjoint to the Fokker-Planck equation with exact integration in the time domain. The numerical solution is very accurate compared to standard methods for computing conditional moments. We used the computed moments in this paper to form a quasi maximum likelihood function for parameter estimation. 
The method is computationally very fast, as the complexity is sublinear. All that is needed is to compute a single matrix exponential to compute all moments, regardless of the number of observations. This makes the method well suited for parameter estimation of large data sets, which was confirmed by Figure~\ref{fig:TimePlot}, in stark contrast to many approximate maximum likelihood methods for which the computational complexity typically is superlinear in the number of observations.


\section*{References}
\bibliographystyle{elsarticle-harv}
\bibliography{references,newreferences}

\end{document}